# Pulse propagation in a hyper-lattice

Joseph Dickey

The Johns Hopkins University, Whiting School of Engineering, 810 Wyman Park Drive, Baltimore Maryland, 21211 USA. Email; Dickey@jhu.edu

**Abstract**

The classical dynamics and pulse propagation are presented for a series of lattice-like structures whose spatial dimensionality ranges from one to four: four representing a hyper lattice. The lattices are connected one-dimensional wave bearing systems of varying lengths and can illuminate some aspects of higher dimension structures. Short pulses are launched at an arbitrary point, reverberate throughout the entire structure, and detected at another point. Some aspects of increasing dimensionality are illustrated with particular emphasis on the transition from three to four spatial dimensions. In a hypothetical four dimension world where only three are observable, the classical conservation laws and causality do not hold. The lack of causality is illustrated at each step in dimensionality by showing the "unexpected" pulse returns from the next higher dimension.

# Pulse propagation in a hyper-lattice

Joseph Dickey

## I. Introduction

In this paper, I calculate the classical dynamics and present the arrival times of reverberant pulses in a series of lattice-like structures whose spatial dimensionality ranges from one to four; four representing a hyper lattice. The structures under consideration are formed from connected one-dimensional wave bearing systems and are fundamentally one-dimensional structures in the sense that propagating waves interact only at junctions. They can, however, illuminate some aspects of structures in higher dimensions. My intent here is to illustrate some of these aspects of increasing dimensionality with particular emphasis on the transition from three to four spatial dimensions. In all examples, the structure is driven harmonically at an arbitrarily chosen point and its response calculated at a second distinct arbitrary point within the same system. The temporal response is then derived from a Fourier transform of the frequency response and is presented as pulse arrival times for the reverberant pulses originating from a short pulse at time zero at the drive point.

Higher dimension dynamics in the literature appears to be confined to string theory, collider physics and cosmology where dimensions outside of the three apparent ones are many orders of magnitude smaller or larger. The trend in these approaches is to *"compact"* some of the extra dimensions which reduces the dimensionality of the calculations. Even the classical dynamics of these structures with highly disparate scales would require a different model than employed here. I make no pretense that the results presented here have any relevance to the world we live in.

There seems to be no prior treatment of classical wave or pulse propagation in higher dimensional structures where the extent in all dimensions is the same order of magnitude. In contrast to this, there are a large number of popular accounts of extra dimensionality, some with scientific merit and many without. A gem among the early treatments is Edwin Abbot's *Flatland* (1), published as a novel in 1884. More recent works by Rucker (2), McMullen (3) and Kaku (4) are examples of popular works written by competent scientists. Brian Greene's book, *The Elegant Universe* (5), is an excellent popular work on the physics of string theory, and the mathematics of higher dimensionality is elegantly outlined in Roger Penrose's book, *The Road to Reality* (6). Regular polytopes with some discussion of higher dimensionality are discussed by Coexter (7).

## II Generating the Lattices

The general rule for generating a lattice of dimension $N+1$ from $N$ is to spatially translate the $N$ lattice in a direction orthogonal to all its existing dimensions. Looking first at $N = 1$, a single wave bearing system shown in Figure 1(A), we get the two dimensional lattice shown in Figure 1(B) through such a translation.[1] Thus, System 1 in Fig. 1(B) is referred to as the *"generator"* for that structure.

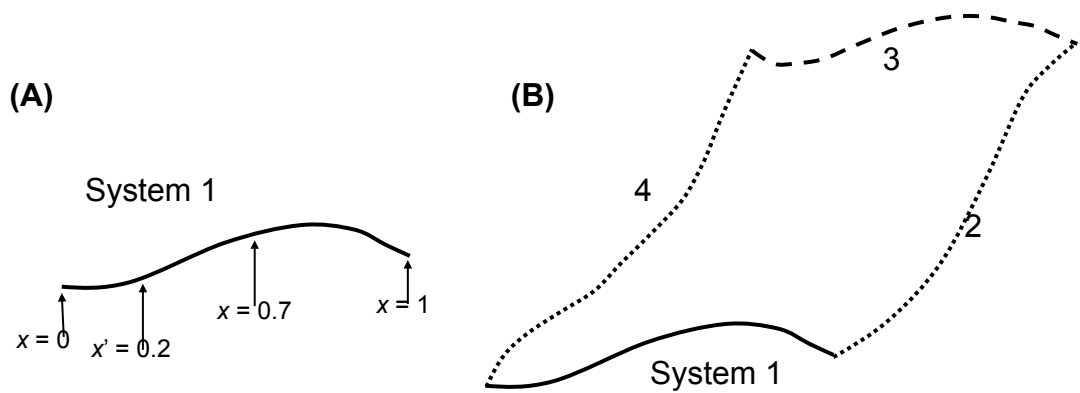

Figure 1. (A) the 1-D and (B) 2-D lattices. The 2-D lattice is generated from the 1-D by translating it in an orthogonal direction: the solid line (System 1) is the *generator*, the dashed line (System 3) is the *image* in the added dimension, and the dotted lines (Systems 2 and 4) are the *connecting* systems.

---

1. The rule works for $N = 0$ also. Clearly, the line is formed by translating a point. I omit this because the dynamics of a point don't exist.

There are two important points to be made here:

1) The lattice depicted in figure 1(B), and any lattice constructed by connecting 1-D systems in this way, is still a 1-D structure; i.e. the only coupling between the 1-D waves of the constituent systems is at the junctions.

2) The translation does not need to preserve lengths; e.g., system 3 in Figure 1(B) is not necessarily the same length as system 1, nor are systems 2 and 4 the same length. In general, for all cases of any dimensionality considered in this paper, the resonant frequencies and impedances of all systems are different.

Applying the rule to the $N = 2$ lattice of Figure 1(B), we get the $N = 3$ lattice shown in Figure 1(C). We are already needing to show only the projection of the $N = 3$ lattice onto the plane of the paper. It can only get worse, as shown by applying the rule once again to get the $N = 4$ hyper-lattice projected onto the plane of Figure 1(D).

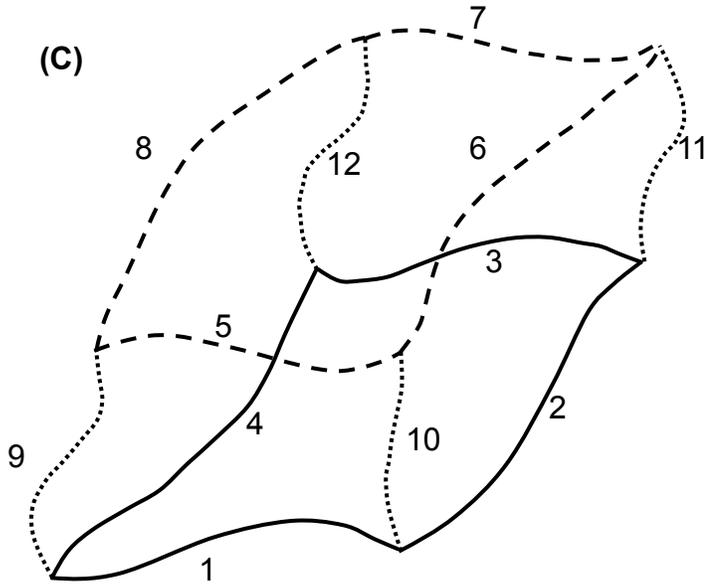

Figure 1. (C) The three dimensional lattice. The solid lines (Systems 1 – 4) form the two dimensional *generator*, the dashed lines (systems 5 – 8) form the *image* in the third dimension, and the dotted lines (systems 9 – 12) are the *connectors*.

**(D)**

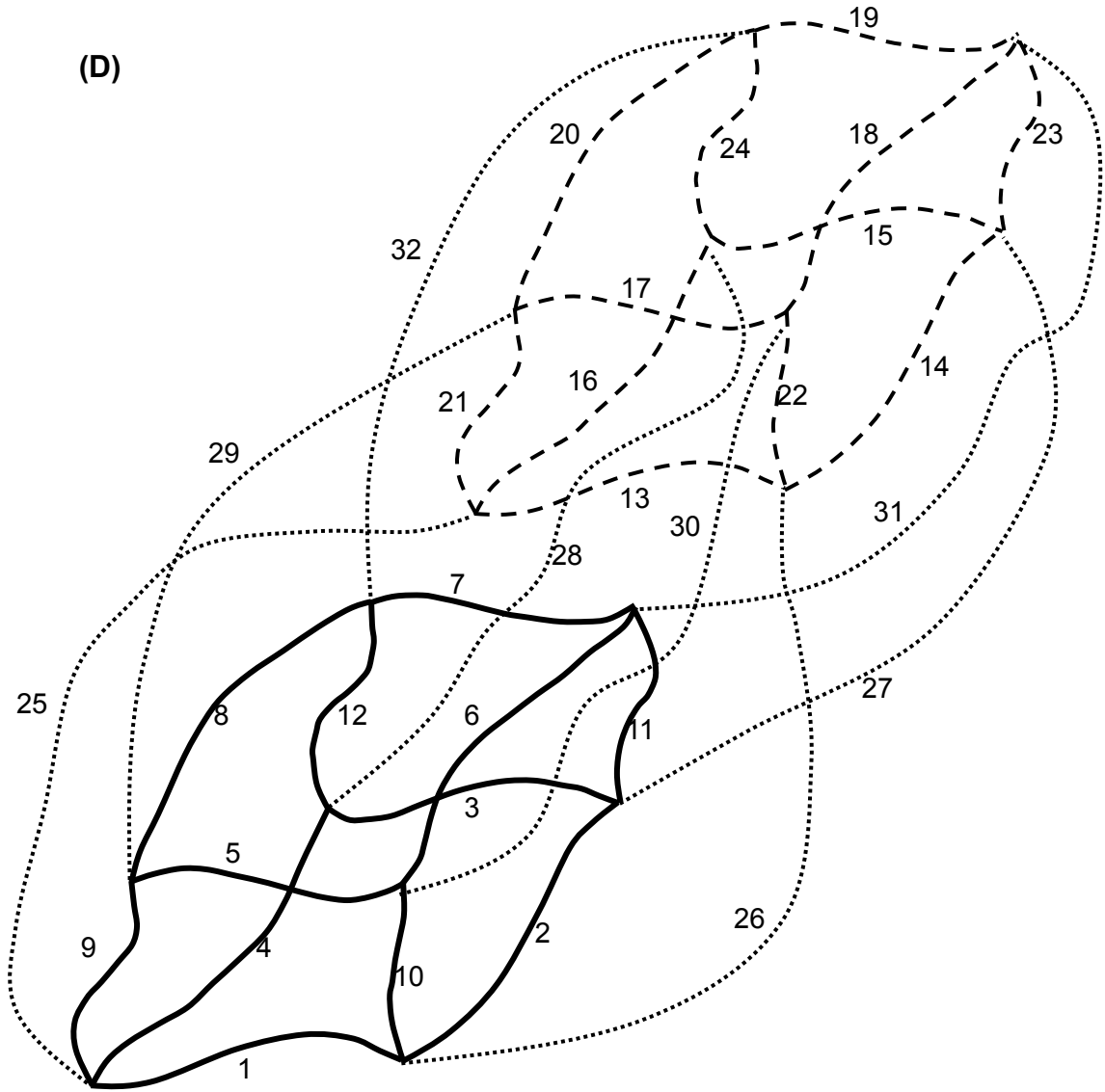

Figure 1. (D) The four dimensional lattice. The solid lines (Systems 1 – 12) form the three dimensional *generator* shown in Figure 1 (C), the dashed lines (Systems 13 – 24) form the *image* in the fourth dimension, and the dotted lines (systems 25 – 32) are *connectors*.

The number of dynamic systems, when progressing with this rule, goes up quickly. If $N$ is the dimensionality and $n_N$ is the number of systems for that dimensionality, we have:

$n_{N+1} = 2n_N + 2^N$. Starting with $N=0$ and $n_0=0$, we get:

| $N$ | $n_N$ |
|---|---|
| 1 | 1 |
| 2 | 4 |
| 3 | 12 |
| 4 | 32. |

Table 1. The number of systems, $n_N$, related to the dimensionality, $N$.

If the $N = 4$ lattice shown in Fig. 1(D) were a regular polytope, i.e. if all systems were the same length, it would be a tesseract, or hypercubic lattice. In addition, if all systems had the same wave speed, $c$, and there were no delays in transmission or reflection at the junctions, it would be a regular lattice. In this case, it would exhibit a highly degenerate dynamics since all systems and combinations would produce harmonics of the fundamental frequency of a single system. There would be a similar degeneracy in the pulse arrival times for pulse reverberations started by a single pulse somewhere in the structure. This would be an uninteresting case to study and so I have considered structures in which the lengths of the constituent systems are in some sense randomized.

It is also important to have some variation in system impedances since, for example, in the case of the 2-D structure of Fig. 1(B), if the impedances of all systems were the same, there would be no reflections at the junctions. Therefore, I have also randomized the impedances.

In all cases, the lengths of the systems and their impedances are normalized by the values for system number one. Thus the randomizing produced distributions about the unit values for length and impedance and a "random" set was adjusted slightly to give unambiguous pulse arrival times that were within the scale of the time interval chosen for calculations, and which afforded reasonable reflection and transmission coefficients at the junctions.

**III Wave Propagation Within the Lattices**

All lattices are comprised of connected 1-D wave-bearing systems. Each system is a *string* in the sense that it supports a single wave type characterized by a wave impedance of type $\rho c$, $\rho$ being the density and *c* the wave speed.

The waves of frequency $\omega$ are plane and within a single system, assuming that the wave meets no discontinuities (e.g. a junction), propagate a response $\psi$ from the coordinates (*x', t'*) to (*x, t*) according to

$$\psi(x,t) = e^{-ik|x-x'|} e^{i\omega(t-t')} \psi(x',t'),$$
$$k = k_0(1-i\eta), \, k_0 = \omega/c \tag{1}$$

where $\eta$ is the propagation loss factor, $t$ is the time and $k$ is the wavenumber. The propagation expressed in Eq.(1) will be different for each system since values of $\eta$ and $c$ will generally be different. We are often interested in a system "frozen in time", i.e. $t = t'$, and can express the spatial propagator as simply $e^{-ik|x-x'|}$.

When the wave does impinge on a junction, it will reflect and transmit with the well known pressure plane wave coefficients (8). Once the parameters for all systems are specified, the propagation in each system and all the reflection and transmission coefficients can be calculated. From these, one can construct the *IRF* for the entire structure (9). This is a *matrix IRF* which relates a response at any point in the structure to an harmonic drive at any other point.

## IV Pulse Propagation Within the Lattices

In the following sections I will present the temporal response for cases where $N = 1$ through 4. The wave speeds for all systems, $c$, are taken to be unity and the densities and lengths are varied to give the requisite variation in impedances and system resonant frequencies respectively. Also for each case, the lattice is driven in system number 1 at the point $x'=0.2$ as measured from the left-end as shown in Fig. 1(A), and with a unit amplitude harmonic localized force of frequency ω. The force response is assessed at the point $x=0.7$ as a function of ω and is calculated by using the above-mentioned *matrix IRF*. The pulse arrival times at *x* resulting from a unit-amplitude short pulse at *x'* are then calculated by performing a *numerical Fourier transform* on the frequency data set.

It is instructive to illustrate the additional pulse arrivals resulting from the *N* to *N*+1 increase in dimensionality. To show this, I subtracted the *N* arrivals from the *N*+1 arrival data by increasing the lengths of the *connecting* systems in the *N*+1 structure. Using *N*=3 as an example and referring to Fig. 1(C), the pulse arrival spectrum for *N*=3 was calculated. A separate calculation was then made wherein the connecting systems, i.e. systems number 9,10,11 and 12 were made long enough that there would be no returns from these systems during the time frame of interest (∞ would have worked just fine) while keeping their characteristic impedances unchanged. This left the terminating characteristic impedances unchanged; and thereby the reflection and transmission coefficients unchanged, and finally the pulse amplitudes of the reverberant pulses within the generating system, unchanged. This process did, however, remove all returns involving the *connecting* systems. The response obtained in this way is termed the "*in vivo*" response, whereas if the connecting impedances were set to zero, the response would be termed the "*in vitro*" response. When the *in vivo* response is subtracted from the total response, one is left with just the returns which depend on the higher dimensionality of the $N=3$ structure.

**A. Pulse Arrival Times for $N = 1$**

The 1-D case is a single system as shown in Fig. 1(A). In order to calculate the frequency response of the 1-D system, we need to specify the following parameters:

1) The length and wave speed of the system. These are both unity for System 1.

2) The terminating impedances. These are zero for both.

3) The propagation loss factor defined in Eq.(1). This is $\eta_1 = 0.003$.

4) Any delays upon reflection at the ends. All reflections and transmissions for this and all structures in this paper are zero.

5) The points, $x'$ and $x$, where the drive is applied and assessed respectively. They are $x'=0.2$ and $x=0.7$, as shown in Fig. 1(A).

I give these values only by way of example, and stress that giving the impedances, loss factors, and lengths for all the systems in the various structures would obfuscate the point of the paper. It should also be noted that parameters 2) and 3) above would make no difference in the pulse arrival times for the structure; they would however affect the rate of decay in the reverberant spectra.

The calculated frequency response is shown in Fig. 2 (A). A numerical Fourier transform was performed on this and the result, shown in Fig. 2 (B), gives the arrival times at the point $x$ of a unit amplitude short pulse which left the point $x'$ at time $t = 0$.

The interpretation is pretty clear. All arrival times in Fig. 2 (B) are associated with the pictographs shown in Fig. 3. The arrivals at $t = 2.9$ and $3.1$ in Fig. 2 (C) are simply the 0.9 and 1.1 arrivals plus an additional round trip.

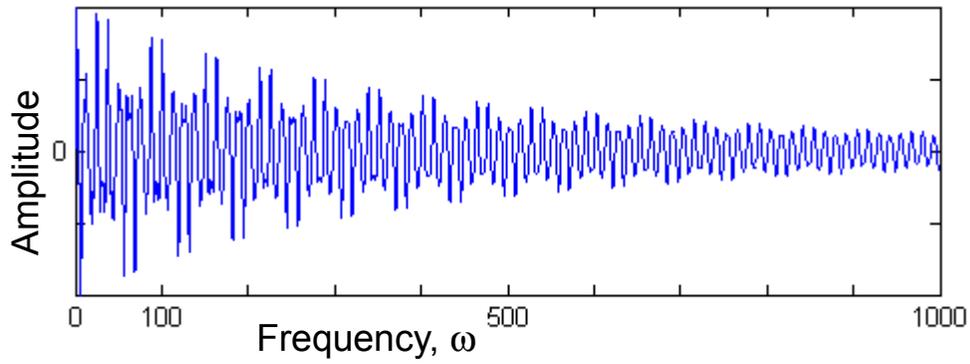

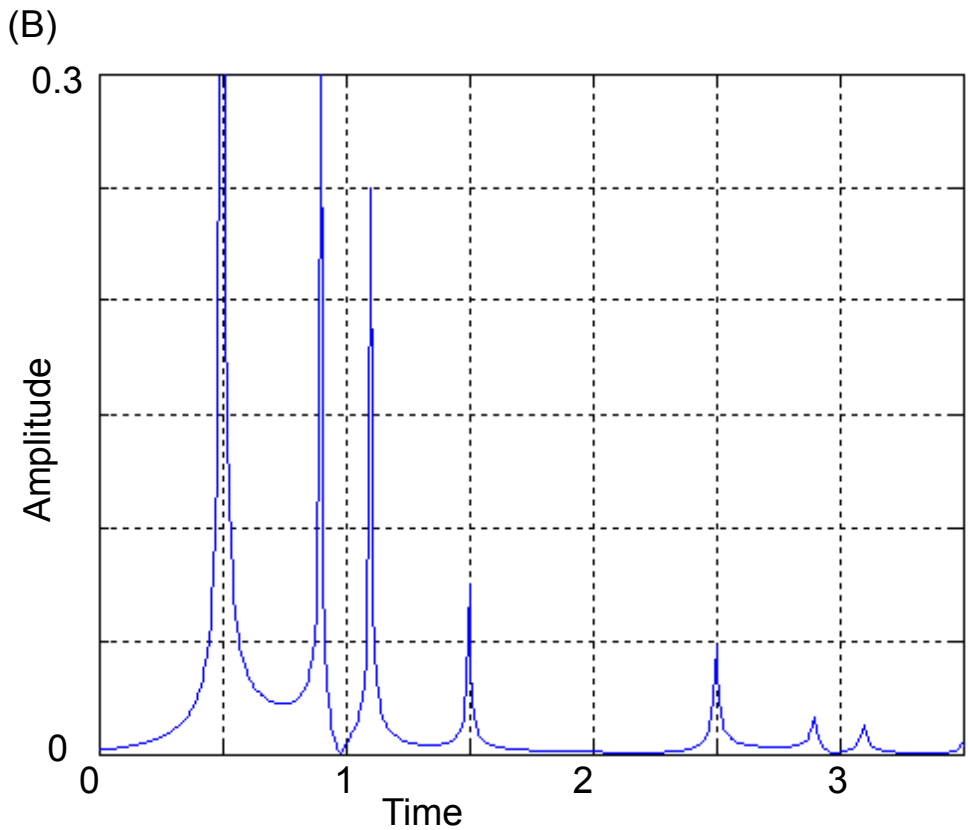

Figure 2. (A) The frequency response of the single 1-D system at the assessment point $x = 0.7$ when driven with a unit-amplitude harmonic drive of frequency $\omega$ at the drive point $x' = 0.2$. (B) The pulse arrivals at the assessment point generated by a unit amplitude short pulse at the drive point at time = 0.

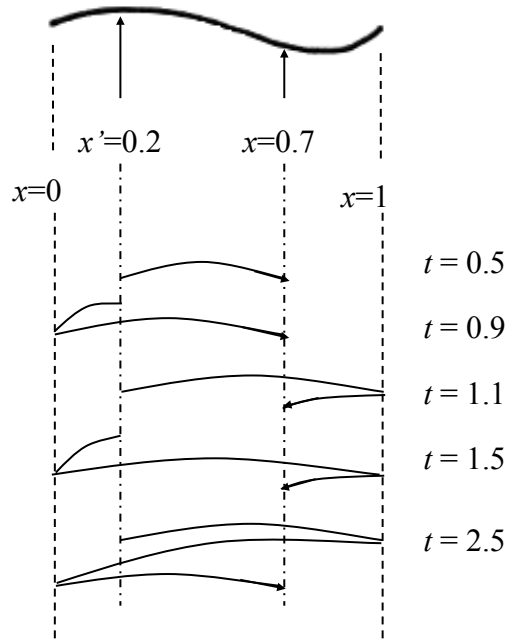

Figure 3. The paths and arrival times for the first five returns at the point x in the single, 1-D, system driven by a short pulse at t=0 at x'.

**B. Pulse Arrival Times for $N = 2$**

The 2-D case is the structure shown in Fig.1(B). System 1 in this Figure is the *generator* and is identical to the single system used in the 1-D case, except that its terminating impedances are different. These impedances are $z_2=1.7$ and $z_4=1.8$ for the right and left hand terminations respectively. Again, the structure is driven and assessed at the same two points in System 1 and the temporal response, as derived from the Fourier transform of the calculated frequency response, is shown in Fig. 4(A). In this figure, I have superposed the *in vivo* response of the single system as a dashed line for comparison.

As expected, there is very little difference between the responses of the 1-D *in vivo* and 2-D structures until such time as the pulse returns involving Systems 2, 3 and 4 arrive; i.e. at t = 2.5. Fig. 4(B) shows the difference between the *in vivo* 1-D system and the 2-D structure responses with an expanded scale. This shows more clearly the pulse arrivals which depend on the additional systems required to make the 2-D structure. Every peak in Fig. 4 (B) can be associated with a path from *x* to *x'* involving at least one of the Systems 2, 3 or 4.

Note that the peaks at t = 4.5 and 4.9 are degenerate in the sense that they are both 1-D and 2-D peaks. This is why they are larger in the 2-D case and show prominently in Fig. 4(B).

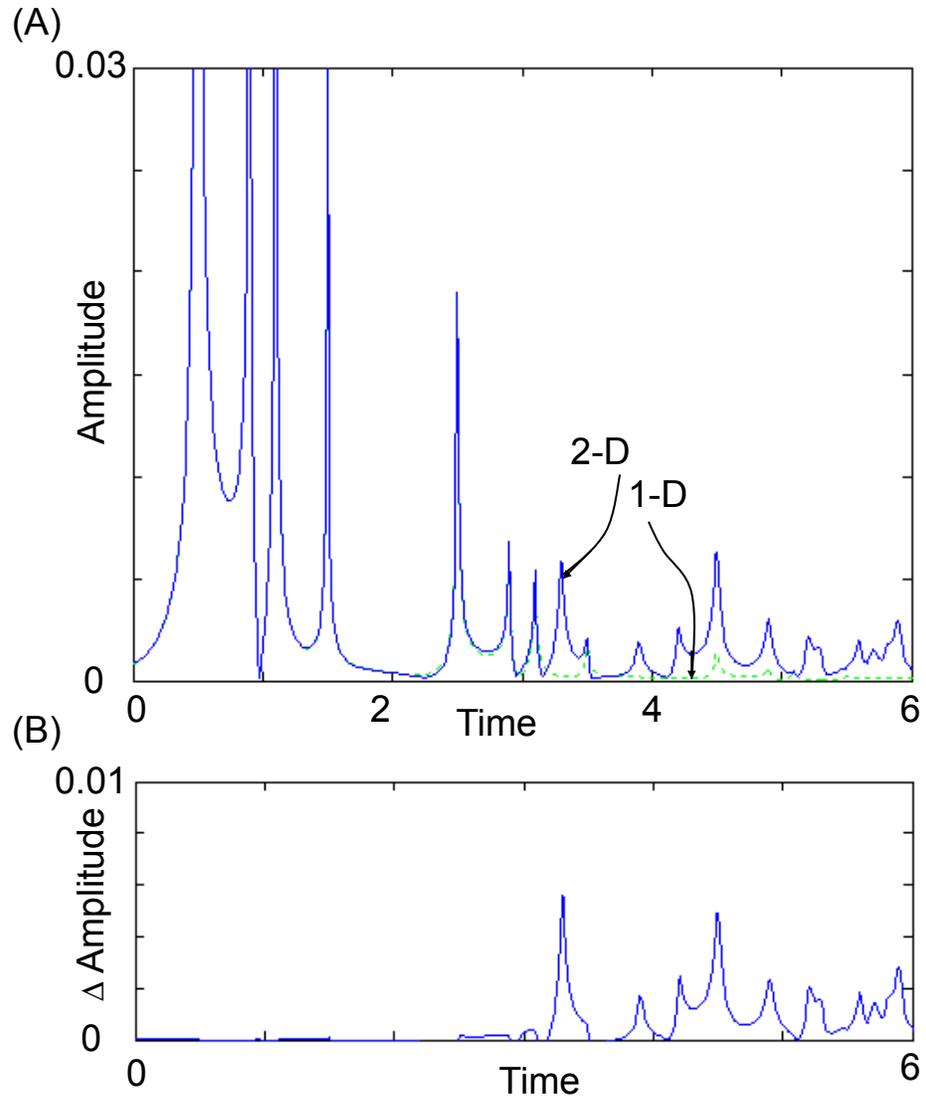

Figure 4. The temporal response of the 2-D structure shown in Fig. 1(b). (A) The response of both the 2-D (solid) and 1-D (dashed). (B)

17   Dickey, Hyper-lattice, arXiv

## C. Pulse Arrival Times for $N = 3$

The 3-D structure is shown in Fig. 1(C). Again, the structure is driven and assessed at the same two points in System 1. The temporal response of the 3-D structure is again calculated from the Fourier transform of the calculated frequency response and shown in Fig. 5(A). The *in vivo* response of the 2-D *generator* structure is now calculated using the additional characteristic impedances of Systems 9, 10, 11 and 12 attached as shown and making these systems long. This *in vivo* response is also shown in Fig. 5(A) as a dashed line. Figure 5(B) shows the pulse returns in the $N=3$ structure which were not in the $N=2$ structure; i.e. the extra response of the structure resulting from the added dimensionality. As before, and as expected, there is very little difference between the *in vivo* and *total* response until such time as the pulse returns involving Systems numbered >2 arrive. There are some degenerate responses, e.g. the one at t = 3.1.

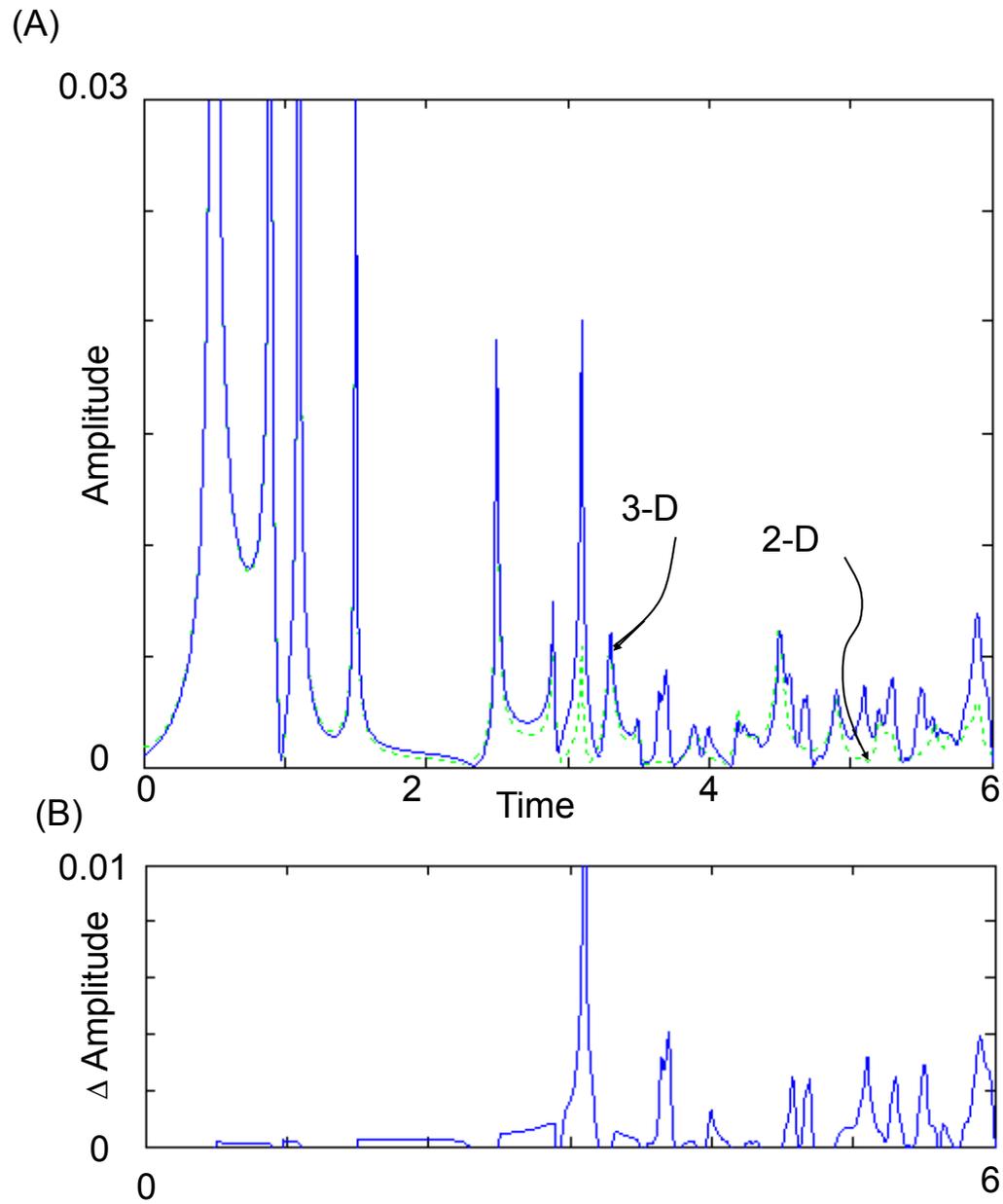

Figure 5. The response of the 3-D structure shown in Fig.1(C). (A) The response of both the 3-D (solid) and 2-D (dashed). (B) The excess of 3-D over 2-D on a magnified scale.

**D. Pulse Arrival Times for *N* = 4**

The 4-D structure is shown in Fig. 1(D). Again, the structure is driven and assessed at the same two points in System 1. The temporal response of the 4-D structure is shown in Fig. 6(A). The *in vivo* response of the 3-D *generator* structure is also shown as a dashed line in this figure. The responses in the 3-D structure resulting from the extra dimensionality are shown in Fig. 6(B).

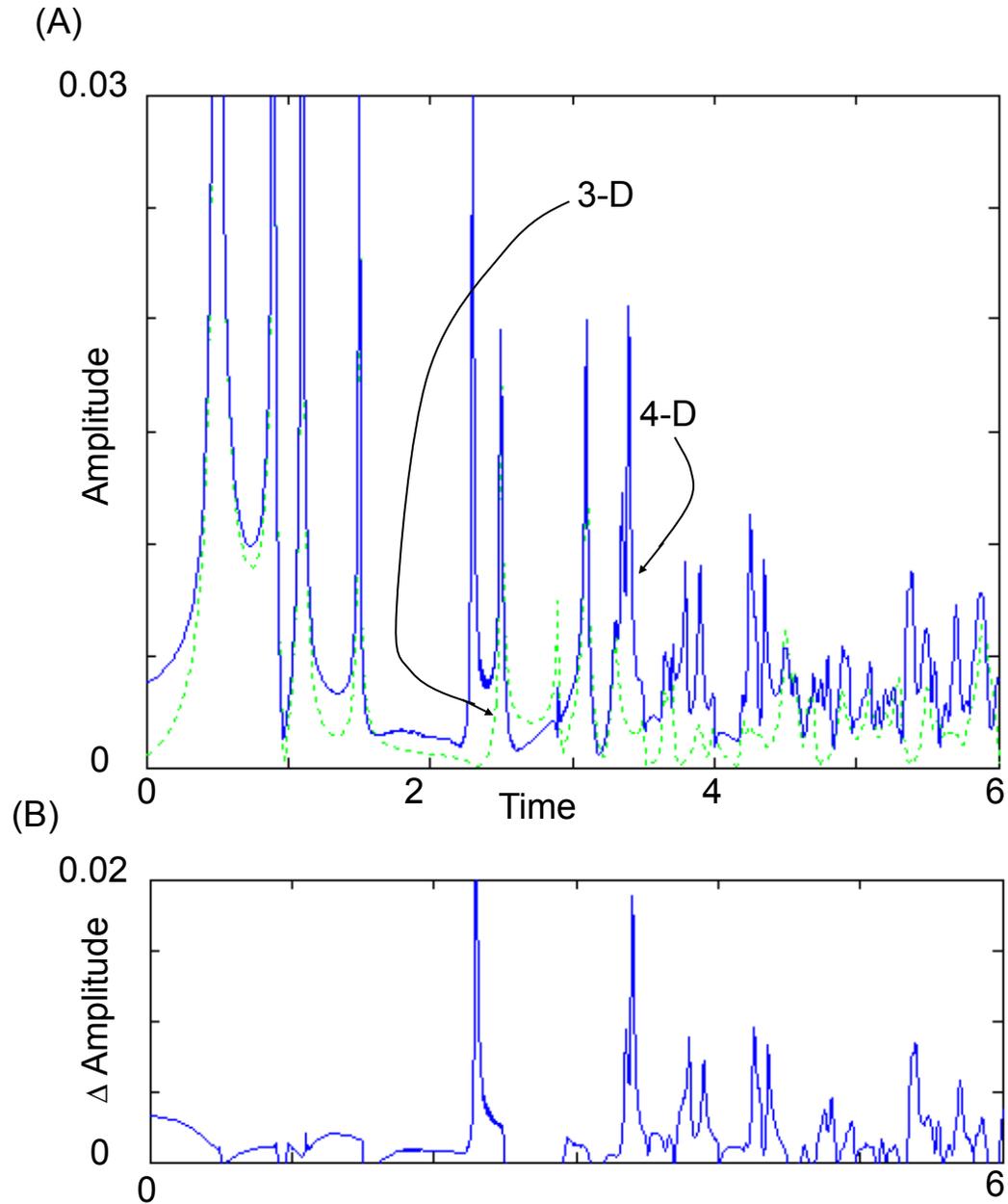

Figure 6. The response of the 4-D structure shown in Fig.1(D). (A) The response of both the 4-D (solid) and 3-D (dashed). (B) The excess of 4-D over 3-D on a magnified scale.

**V. Some Afterthoughts**

A major strength of the present work is that the calculations in the frequency domain are exact (at least within the approximations of linear wave theory). The discrete Fourier transforms used to get pulse arrival times introduce a degree of approximation; however, the arrival times are clear, unambiguous and closely match the expected arrival times.

A major shortcoming is that the structures are fundamentally 1-D. For a truly *N*-D, with *N>1*, structure, the propagating pulses interact continuously with the other dimensions as they propagate. In connected 1-D systems as employed here, they interact only at junctions. This then relates to another shortcoming: I considered only a single wave type. Even a 1-D string has both longitudinal and transverse (compression and flexural) wave types and the transverse waves have two polarizations. In a 3-D world, this is three waves, one for each dimension. For an *N*-D world, there would be *N* wave types all propagating in the same direction. We wouldn't see any more than three and, at least in our macro-world, any extra dimensions couldn't disturb energy conservation or causality; at least within our experience and ability to measure the conserved quantaties. That is, there can be no significant losses in the extra dimensions.

One can posit that extra dimensions, if they exist, are fundamentally different from our three observable ones. For example, the impedance which these waves see in the extra dimensions may be purely reactive with no energy flow except for possible tunneling to observed dimensions if the connecting dimensions were small and compact.

I believe it is possible to calculate the frequency domain *matrix IRF* for small and compact dimensions using classical waves, but I question the utility of this. Also, the Fourier transform needed to get the pulse propagation in structures greater than 1-D with highly disparate dimensions looks problematical; however, it may be possible to calculate the *matrix IRF* directly in the time domain (10).